# Fast-light Enhanced Brillouin Laser Based Active Fiber Optics Sensor for Simultaneous Measurement of Rotation and Acceleration

**Minchuan Zhou,[1] Zifan Zhou,[2] Mohamed Fouda,[2] Nicholas Condon,[3] Jacob Scheuer,[4] and Selim M. Shahriar[1,2,*]**

[1]Department of Physics and Astronomy, Northwestern University, Evanston, IL 60208,USA
[2]Department of EECS, Northwestern University, Evanston, IL 60208,USA
[3]Digital Optics Technologies, Inc., Rolling Meadows, IL 60008,USA
[4]School of Electrical Engineering, Tel-Aviv University, Ramat Aviv, Tel-Aviv 69978, Israel
*Email: shahriar@northwestern.edu

**Abstract** We have developed a conceptual design for an Active Fast Light Fiber Optic Sensor (AFLIFOS) that can perform simultaneously or separately as a gyroscope (differential mode effect) and a sensor for acceleration, strain, and other common mode effects. Two Brillouin lasers in opposite directions and separated in frequency by several free spectral ranges are used for this sensor. By coupling two auxiliary resonators to the primary fiber resonator, we produce superluminal effects for two laser modes. We develop a detailed theoretical model for optimizing the design of the AFLIFOS, and show that the enhancement factor of the sensitivity is ~187 and ~−187, respectively for the two Brillouin lasers under the optimized condition, when the effective change in perimeter of the primary fiber resonator is 0.1nm, corresponding to a rotation rate of 0.4 deg/sec for a ring resonator with radius 1m. It may be possible to get much higher enhancement by adjusting the parameters such as the perimeters and the coupling coefficients.

## I. INTRODUCTION

Optical interferometers and resonators are commonly used for precision measurement. These devices can measure accurately the phase shifts of light due to parameters of interest, including rotation, acceleration, strain and temperature. Recently, much theoretical and experimental work [1,2,3,4,5,6,7,8,9,10,11,12,13,14,15,16] has been carried out for enhancing the sensitivity of such devices using the superluminal effect. These studies considered both passive [1,3,4,5,6,7,9,10,12,13,14,15] and active [1,2,3,4,5,6,8,9,11,12] versions of the resonators. Furthermore, different physical mechanisms have been considered, in both active and passive cases, for realizing the anomalous dispersion that produces the superluminal effect. Here, we consider the case of an active resonator, for which the superluminal effect is produced via coupling to other resonators [17]. Specifically, we describe a system called Active Fast Light

Fiber Optic Sensor (AFLIFOS). It consists of a primary fiber resonator that is coupled to two auxiliary resonators for creating the superluminal effects for two counter-propagating Brillouin laser modes.

The rest of the paper is organized as follows. In Sec. II, we introduce the realization of the fast light effect using additional cavity resonators. In Sec. III, we describe the scheme of the AFLIFOS. In Sec. IV, we theoretically model the AFLIFOS and show the results of the sensitivity enhancement factor under optimized condition. Finally in Sec. V, we describe the design for stabilizing the perimeters of the CPEs using master lasers, and the readout section for processing the output signal.

## II. CAVITY PHASE ELEMENT

The basic concept of the AFLIFOS is schematically illustrated in Figure 1(a). Two non-degenerate pump beams (P1 and P2) are sent into a fiber resonator and two counter-propagating Brillouin lasers (B1 and B2) are generated opposite to the propagating directions of the respective pump beams. Two additional rings are coupled to the primary ring resonator in order to produce anomalous dispersion. The details of the scheme will be discussed later in Sec. III. In this section, we discuss in detail the functionality of each of these auxiliary resonators, focusing on the section in Figure 1(a) that is enclosed with dashed lines. This section is illustrated in greater detail in Figure 1(b). In what follows, we will refer to this section as a cavity phase element (CPE).

As illustrated in Figure 1(b), a CPE consists of a fiber ring resonator coupled to the primary resonator as well as another segment of fiber, and can be used to produce a negative dispersion. Here, $a_i$ and $b_i$ are complex amplitudes of the fields. We denote the intensity coupling coefficients as $k_i$ ($i=1,2$) for the two couplers. Assume that each coupler is internally lossless. Using the transfer matrix method, we calculate the effective transmissivity and reflectivity of the CPE as

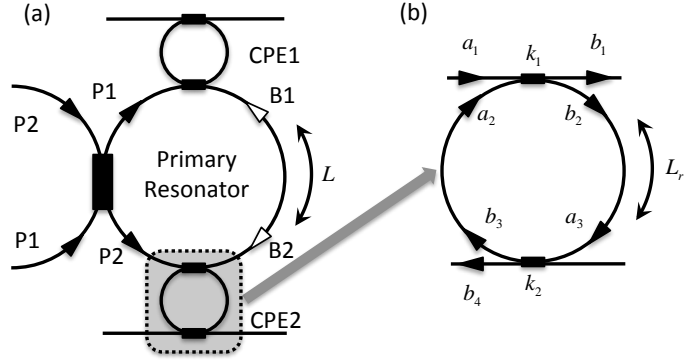

Figure 1 (a) Schematic illustration of the basic concept of AFLIFOS; (b) Cavity phase element.

$$t_{CPE} \equiv \frac{b_1}{a_1} = \frac{\sqrt{1-k_1} - \sqrt{1-k_2}e^{i\phi}}{1 - e^{i\phi}\sqrt{1-k_1}\sqrt{1-k_2}} \equiv |t_{CPE}|e^{i\theta_{CPE}} \;, \qquad (1)$$

$$r_{CPE} \equiv \frac{b_4}{a_1} = \frac{-e^{i\phi/2}\sqrt{k_1}\sqrt{k_2}}{1 - e^{i\phi}\sqrt{1-k_1}\sqrt{1-k_2}} \;, \qquad (2)$$

where $\phi = n_0 \omega L_r / c$ is the phase shift in the ring, $n_0$ is the effective index of refraction of the fiber, $L_r$ is the perimeter of the ring resonator, and $\theta_{CPE}$ is the net phase shift imparted by the CPE. The value of $\theta_{CPE}$ follows immediately from Eq. (1):

$$\theta_{CPE} = \tan^{-1}\left[-\frac{k_1\sqrt{1-k_2}\sin(\phi)}{\sqrt{1-k_1}(2-k_2) - \sqrt{1-k_2}(2-k_1)\cos(\phi)}\right], \qquad (3)$$

We plot in Figure **2** the amplitude $|t_{CPE}|$ and the phase $\theta_{CPE}$ as a function of detuning, $\Delta/(2\pi)$, from the resonant frequency for the CPE, where we can see that the CPE produces a negative dispersion. At the resonant frequency of the ring resonator, which satisfies the condition that $\phi = 2\pi l$ (where $l$ is an integer), we get maximal negative slope of the phase.

As noted above, the AFLIFOS architecture employs two non-degenerate, counter-propagating Brillouin pumps. In Section IV, we will consider in detail the behavior of the complete system. Here, as an illustration, we consider the case where a single CPE and a single Brillouin pump are present in the system.

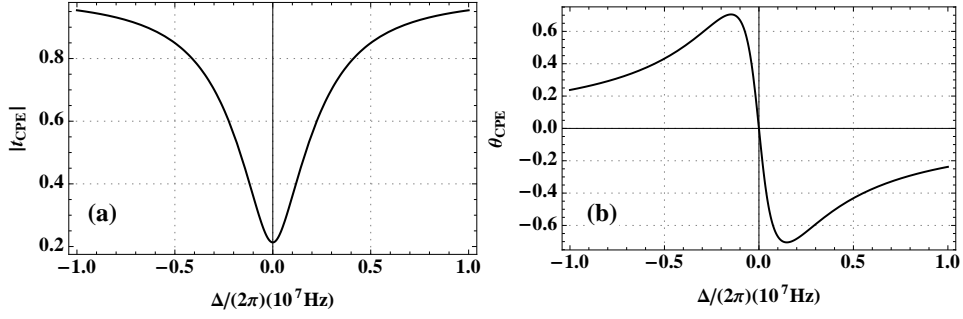

Figure 2 Plots of (a) $|t_{CPE}|$ and (b) $\theta_{CPE}$ as a function of detuning $\Delta/(2\pi)$ for the CPE.

The primary resonator has a round-trip length $L$ [see Figure 1(a)]. If $\omega_{res}$ is a frequency that is resonant in the primary resonator, then it follows that

$$\omega_{res} n_0 L / c + \theta_{CPE}(\omega_{res}) + \theta_{Br}(\omega_{res}) = 2\pi K, \tag{4}$$

where $K$ is an integer, and $\theta_{Br}$ represents the phase shift from the dispersion induced by the Brillouin pump. The real and imaginary parts of the susceptibility induced by the Brillouin pump can be expressed as:

$$\chi' = \frac{2G(\omega - \omega_0)\Gamma}{2\alpha I + \Gamma^2 + 4(\omega - \omega_0)^2}, \tag{5}$$

$$\chi'' = -\frac{G\Gamma^2}{2\alpha I + \Gamma^2 + 4(\omega - \omega_0)^2}, \tag{6}$$

where $G$, $\omega_0$, $\Gamma$, and $I$ are, respectively, the gain parameter, the peak frequency of the Brillouin gain, the linewidth (of the Brillouin gain), and the probe intensity per unit area. The parameter $\alpha$ is such that $\alpha I$ has the dimension of $\sec^{-2}$, and $\sqrt{\alpha I}$ represents power broadening of the Brillouin gain process. The lasing condition requires that the gain compensates the overall roundtrip losses. Suppose the quality factor of the primary resonator is $Q$ in the absence of the Brillouin pump. Then the roundtrip transmissivity of the field amplitude in the cavity is

$$t_{cav} = \exp[-k n_0 L / (2Q)], \tag{7}$$

Taking into account the transmissivity $|t_{CPE}|$ of the CPE, we can write (for gain>loss at the onset)

$$t_{cav} |t_{CPE}| \exp[-\chi''(I) n_0 k L / 2] = 1. \tag{8}$$

Solving Eq. (8) we can get the value of $I$ as a function of frequency, from which we can determine $\theta_{Br}$ after lasing, using

$$\theta_{Br} = \chi' k n_0 L / 2 . \tag{9}$$

The white light cavity (WLC) condition (which corresponds to maximum superluminal enhancement) requires that the phase accumulated in a round-trip remains $2\pi K$ regardless of the frequency for a range of frequencies near the resonant frequency. For a frequency that is $\Delta\omega$ away from $\omega_{res}$, we then have, for the WLC condition:

$$(\omega_{res} + \Delta\omega) n_0 L / c + \theta_{CPE}(\omega_{res} + \Delta\omega) + \theta_{Br}(\omega_{res} + \Delta\omega) = 2\pi K . \tag{10}$$

From Eqs. (4) and (10), using the first order approximation in $\Delta\omega$ that

$$\left.\frac{d\theta}{d\omega}\right|_{\omega_{res}} = \frac{\theta(\omega_{res} + \Delta\omega) - \theta(\omega_{res})}{\Delta\omega}, \tag{11}$$

we get

$$\left.\frac{d\theta}{d\omega}\right|_{\omega_{res}} + \left.\frac{d\theta_{Br}}{d\omega}\right|_{\omega_{res}} = -\frac{n_0}{c} L , \tag{12}$$

We take the resonant frequency $\omega_{res}$ to be the same as the resonant frequency of the auxiliary resonator in the CPE. For a given set of $L$, $L_r$, and $\omega_{res}$, the left-hand side of Eq. (12) depends on the values of $k_1$ and $k_2$. In our simulation, we fix $k_1$ as 0.1, and determine $k_2$ from this equation.

### III. DESCRIPTION OF AFLIFOS

Figure 3 shows a schematic diagram of the ring resonator sensor embodying the AFLIFOS. The source laser is locked to one of the optical resonances in an acetylene vapor cell using saturated absorption spectroscopy [18, 19, 20]. For specificity, we consider the laser to be locked to the R(5) line at 1530.24nm in the $v_1 + v_3$ bands of $^{13}C_2H_2$, whose uncertainty can be as small as 0.7kHz [21, 22]. In our simulation, we use for simplicity the wavelength of 1530nm, rather than 1530.24nm. This choice does not affect the results for the enhancement factor significantly. The output of the locked laser is passed through an isolator, and split by a beam splitter (BS1) into two equal-powered beams. One of these beams is again split into two equal-powered beams P1 and P2 by another beam splitter (BS2), which are frequency shifted by two separate acoustic-optic modulators (AOM-1A and AOM-1B, respectively) after passing through two separate circulators (Circulator1 and Circulator2, respectively). The resulting beams, P1 and P2, are coupled to the primary fiber ring resonator in opposite directions. The frequencies of AOM-1A and AOM-1B differ by an amount that is $m$ times the free spectral range of the primary resonator, where $m \geq 1$ is an integer. Thus, the two laser modes are non-degenerate, which enables determination of the sign of rotation and eliminates the lock-in problem [23,24]. Two Brillouin lasers, B1 and B2, are produced along directions opposite to those of the respective pumps. The outputs from port 3 of Circulator1 and Circulator2 are each sent through a Fabry-Perot cavity, which blocks out the pump, and then sent to the readout section. The fiber resonator is coupled to cavity phase elements CPE1 and CPE2 in order to produce superluminal effects.

The other beam produced by BS1 is split into two equal-powered beams Pr1 and Pr2 by another beam splitter (BS3). These are frequency shifted by AOM-2 and AOM-3, respectively. Each of these "master" beams (Pr1 and Pr2) is sent to the side fiber in the corresponding CPE and the output from the fiber is detected and used as a feedback signal to stabilize the perimeter of the CPE with the use of PZTs. Here, for simplicity, we have shown CPE1 and CPE2 to be circular. In practice, the auxiliary cavities are configured in a figure-eight shape in order to eliminate sensitivity of the resonances in these cavities to the Sagnac effect.

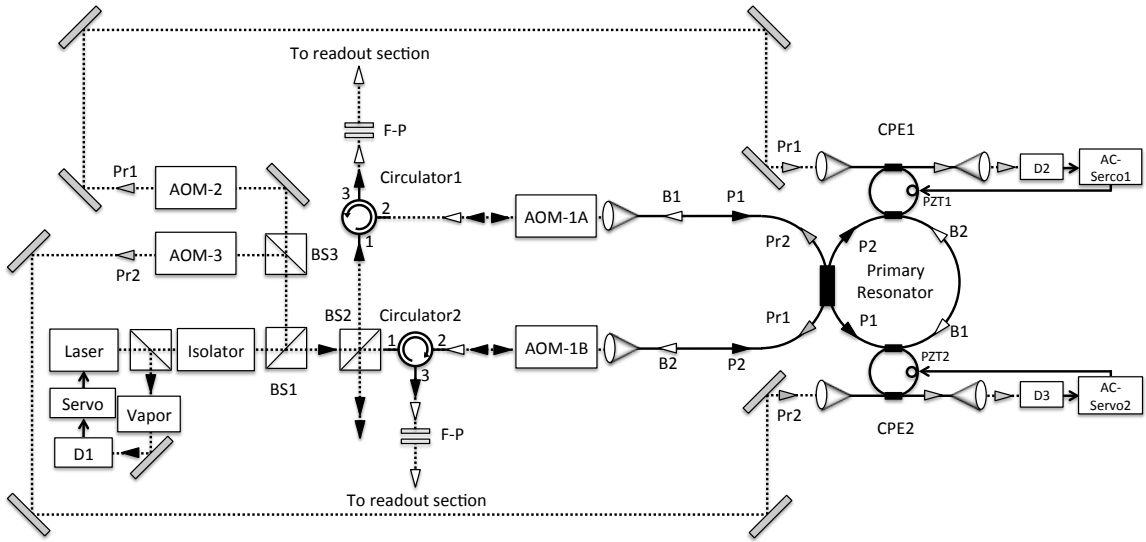

Figure 3 Schematic diagram of the ring resonator sensor.

The ring resonator system can be used to detect differential mode effects, e.g. a rotation rate $\Omega$ orthogonal to its plane. Effectively, the rotation of the resonator can be viewed as a change, $\delta L$, in the resonator perimeter for B1, and an opposite change $-\delta L$ for B2 [5]. With these changes in the perimeter $L$, the frequency changes for B1 and B2 are, respectively, $\delta\omega_1$ and $\delta\omega_2$, with [25]

$$\delta\omega_2 = \omega\delta L/L = -\delta\omega_1. \tag{13}$$

Therefore, the beat frequency of B1 and B2 is $\delta\omega = \delta\omega_2 - \delta\omega_1 = 2\omega\delta L/L$. For rotation sensing, we get, according to the Sagnac effect, that

$$\delta\omega = \frac{\omega}{cn_0}\frac{4\Omega A}{L}, \tag{14}$$

where $\Omega$ is the rotation rate, $A$ is the area enclosed by the resonator, $n_0$ is the index of refraction for the fiber, and $c$ is the vacuum speed of light. We can thus see that detecting changes in the perimeter of the same magnitude $\delta L$ but different signs for two counter-propagating lights is equivalent to detecting a rotation rate of

$$\Omega = \frac{cn_0\delta L}{2A}. \tag{15}$$

For example, for a ring resonator with radius 1m, a change in perimeter of 0.1nm corresponds to a rotation rate of 0.4deg/sec.

Besides the application as a gyroscope, this system can also be used as a sensor for detecting the change in the perimeter of the primary ring resonator. This will correspond to sensing common mode effects such as acceleration and strain. For example, the sensitivity to acceleration can be produced by wrapping a part of the primary resonator around a spring-loaded structure whose dimension will change when subjected to acceleration; a conceptual schematic of such a system is shown in Figure 4. Some length of the primary Brillouin fiber laser cavity is wrapped around a pair of hemicylindrical mandrels, one of which is fixed and one of which can translate in one dimension, separated by a stiff spring. When acceleration is applied along the sensing axis (the dimension in which the free mandrel can translate), the load on the spring, and thus the strain on the fiber, will change, allowing for acceleration sensing. When the acceleration sensing element is added, the common mode signal will now correspond to the acceleration along the sensing axis.

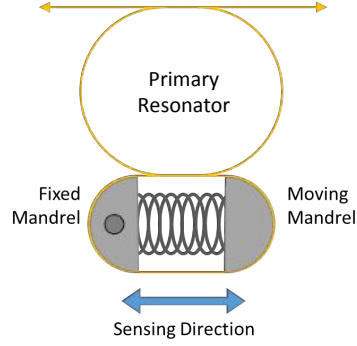

Figure 4 Conceptual sketch of an acceleration/strain transducer for acceleration sensing in an AFLIFOS. Here, for simplicity, we have not shown the two CPE's.

In this design, two different types of beat signals are to be measured: (a) the beat between the two superluminal lasers, and (b) the beat between each superluminal laser and the master beam to which the corresponding CPE is stabilized. These signals can be processed to determine the differential effects (e.g., rotation) and common mode effects (such as acceleration and strain), the details of which are discussed in Sec. V.

## IV. SENSITIVITY ENHANCEMENT

To achieve the optimal condition for sensing, the resonance frequencies of CPE1 and the B1 mode of the primary resonator are both tuned to be at the peak frequency of the Brillouin gain for B1, denoted as $\omega_{01}$. Similarly, the resonance frequencies of CPE2 and the B2 mode of the primary resonator are both tuned to be at the peak frequency of the Brillouin gain for B2, denoted by $\omega_{02}$. It should be noted that the resonance condition for each of the two lasing modes (B1 and B2) are affected by both phase elements (CPE1 and CPE2). We consider the case where $\omega_{01}$ and $\omega_{02}$ are separated by around 12 times the free spectral range (FSR) of the primary ring resonator. The pump (P1/P2) induces Brillouin absorption (gain) for a counter-propagating probe beam having a frequency which is higher (lower) than that of the pump by the Brillouin frequency shift $\omega_B = 10.867\text{GHz}$ [4]. The WLC condition is then determined by the following set of equations

$$\omega_{01} n_0 L / c + \theta_{CPE1}(\omega_{01}) + \theta_{CPE2}(\omega_{01}) + \theta_{Br1}(\omega_{01}) + \theta_{Ab1}(\omega_{01}) = 2\pi M , \qquad (16)$$

$$(\omega_{01} + \Delta\omega) n_0 L / c + \theta_{CPE1}(\omega_{01} + \Delta\omega) + \theta_{CPE2}(\omega_{01} + \Delta\omega) + \theta_{Br1}(\omega_{01} + \Delta\omega) + \theta_{Ab1}(\omega_{01} + \Delta\omega) = 2\pi M , \quad (17)$$

$$\omega_{02}n_0L/c+\theta_{CPE1}(\omega_{02})+\theta_{CPE2}(\omega_{02})+\theta_{Br2}(\omega_{02})+\theta_{Ab2}(\omega_{02})=2\pi N \,, \tag{18}$$

$$(\omega_{02}+\Delta\omega)n_0L/c+\theta_{CPE1}(\omega_{02}+\Delta\omega)+\theta_{CPE2}(\omega_{02}+\Delta\omega)+\theta_{Br2}(\omega_{02}+\Delta\omega)+\theta_{Ab2}(\omega_{02}+\Delta\omega)=2\pi N \,, \tag{19}$$

Here $\theta_{CPE1}$ ($\theta_{CPE2}$) is the phase shift from CPE1 (CPE2), $\theta_{Br1}$ ($\theta_{Br2}$) represents the phase shift from the Brillouin gain induced by the Brillouin pumps P1 (P2), $\theta_{Ab1}$ ($\theta_{Ab2}$) represents the phase shift from the Brillouin absorption induced by the Brillouin pump P1 (P2), and $M$ and $N$ are some integers. Notice that only the field that is counter-propagating to the Brillouin pump experiences the corresponding Brillouin gain/absorption.

We assume that $n_0 = 1.45$ is the same for all fiber media, and choose the perimeter of the primary resonator to be $L = L^{(0)} + \Delta L$, where $L^{(0)} = 10.5\text{m}$. To attain resonant pumping and resonant Brillouin lasing simultaneously, we need to scan $\Delta L$ so that while Brillouin lasers (B1 and B2) are on resonance, the pumps P1 and P2 are also close to resonance. The frequency of P1 (P2) is $\omega_{P1} = \omega_{01} + \omega_B$ ($\omega_{P2} = \omega_{02} + \omega_B$), which is assumed to be $\Delta\omega_{P1}$ ($\Delta\omega_{P2}$) away from the nearest cavity mode. Therefore, we find the following conditions:

$$\begin{aligned}(\omega_{P1}+\Delta\omega_{P1})n_0L/c+\theta_{CPE1}(\omega_{P1}+\Delta\omega_{P1})+\theta_{CPE2}(\omega_{P1}+\Delta\omega_{P1})\\+\theta_{Br2}(\omega_{P1}+\Delta\omega_{P1})+\theta_{Ab2}(\omega_{P1}+\Delta\omega_{P1})=2\pi J\end{aligned} \,, \tag{20}$$

$$\begin{aligned}(\omega_{P2}+\Delta\omega_{P2})n_0L/c+\theta_{CPE1}(\omega_{P2}+\Delta\omega_{P2})+\theta_{CPE2}(\omega_{P2}+\Delta\omega_{P2})\\+\theta_{Br1}(\omega_{P2}+\Delta\omega_{P2})+\theta_{Ab1}(\omega_{P2}+\Delta\omega_{P2})=2\pi J\end{aligned} \,, \tag{21}$$

where $J$ is an integer.

We choose the perimeters of CPEs to be $L_{CPE1} = L_{CPE1}^{(0)} + \Delta L_{CPE1}$ and $L_{CPE2} = L_{CPE2}^{(0)} + \Delta L_{CPE2}$, where $L_{CPE1}^{(0)}$ and $L_{CPE2}^{(0)}$ are each fixed to be 0.82m. For CPE1, we define $k_{11}$ and $k_{21}$ to be the intensity coupling coefficients for the interface with the primary resonator and the external fiber, respectively. The corresponding coefficients for CPE2 are defined as $k_{12}$ and $k_{22}$. With $k_{11} = k_{12} = 0.1$, we scan $\Delta L$, $\Delta L_{CPE1}$, $\Delta L_{CPE2}$, $k_{21}$ and $k_{22}$ to obtain the optimal operating condition. The optimal values are found to be $\Delta L \approx 1.14 cm$, $\Delta L_{CPE1} \approx \Delta L_{CPE2} \approx 0.7807\text{mm}$, and $k_{21} \approx k_{22} \approx 0.1506$. In this case, the frequency difference between the pumps and the nearest resonances are $\Delta\omega_{P1}/(2\pi) \approx 7.5\text{kHz}$ and $\Delta\omega_{P2}/(2\pi) \approx -7.9\text{kHz}$, which are more than 58 times smaller compared to the linewidth of the primary resonator of 0.46MHz. The amount of frequency shifts for the various beams are as follows: ~615.5MHz for AOM-1A, ~850.8MHz for AOM-1B, ~335.3MHz for AOM-2, and ~597.8MHz for AOM-3.

When the change in the perimeter for B1 is $\delta L$ while that for B2 is $-\delta L$, the change in resonant frequency for B1, denoted by $\Delta \omega_{c1}$, and that for B2, denoted by $\Delta \omega_{c2}$, are determined by

$$\vartheta_{rt1} = (\omega_{01} + \delta \omega_{B1}) n_0 (L + \delta L)/c + \theta_1(\omega_{01} + \delta \omega_{B1}) + \theta_2(\omega_{01} + \delta \omega_{B1}) + \theta_{Br1}(\omega_{01} + \delta \omega_{B1}), \quad (22)$$

$$\vartheta_{rt2} = (\omega_{02} + \delta \omega_{B2}) n_0 (L - \delta L)/c + \theta_1(\omega_{02} + \delta \omega_{B2}) + \theta_2(\omega_{02} + \delta \omega_{B2}) + \theta_{Br2}(\omega_{02} + \delta \omega_{B2}), \quad (23)$$

$$\vartheta_{rt1} = 2\pi M, \quad \vartheta_{rt2} = 2\pi N, \quad (24)$$

where $M$ and $N$ are the same as in Eqs. (16)-(19). We plot in Figure 5 the total roundtrip phase $\vartheta_{rt1}$ for B1 around $\omega_{01}$ and $\vartheta_{rt2}$ for B2 around $\omega_{02}$ when $\delta L$ is 0.1nm. We denote the change in resonant frequency for B1 by $\Delta \omega_{c1}$ and that for B2 by $\Delta \omega_{c2}$. For the condition in Figure 5, the enhancement factor for B1 ($\xi_1 = \delta \omega_{B1}/\delta \omega_0$) is ~184 and that for B2 ($\xi_2 = \delta \omega_{B2}/\delta \omega_0$) is ~ −184. Notice that $\delta \omega_0$ is the change in resonant frequency when no CPEs exist in the system. By detecting the beat frequency (Details of the detection scheme is described in Sec. V), we can get an enhancement of ~184. Of course, akin to what was done in Ref. 17, by adjusting the parameters such as the perimeters and the coupling coefficients, it is possible to get much higher enhancement.

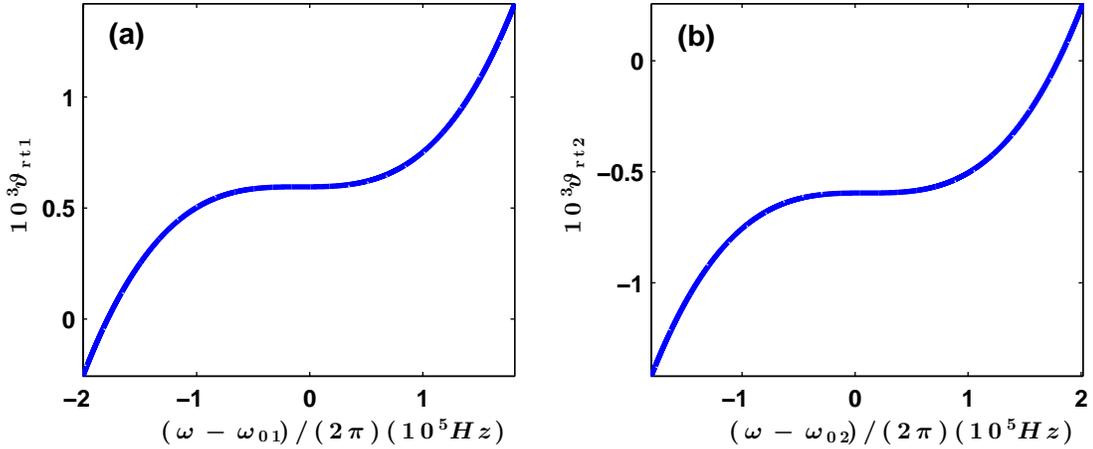

Figure 5 (a) Plot of roundtrip phase $\vartheta_{rt1}$ versus frequency for B1 around $\omega_{01}$ with $\delta L > 0$; (b) plot of roundtrip phase $\vartheta_{rt2}$ for B2 around $\omega_{02}$ with $\delta L < 0$.

When the system is used for common mode sensing, the changes in the perimeters $\delta L$ are in the same direction for B1 and B2. The system can be operated in the same condition as in Figure 5, and the enhancement factor is ~184 for both B1 and B2 when $\delta L = 0.1 nm$.

In Figure **6**, we show that the enhancement factor $\xi_1 \cong \xi_2 \equiv \xi$ increases with smaller change in the perimeters $\delta L$. The equivalent rotation rate $\Omega$ determined by Eq. (13) is also shown.

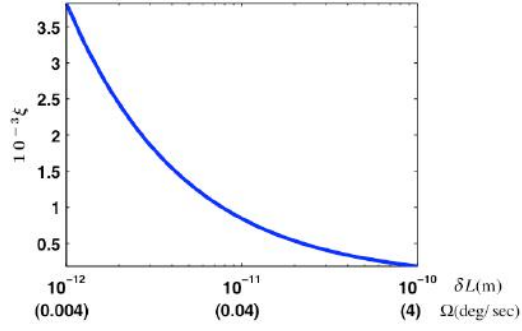

Figure 6 Plot of the enhancement factor $\xi$ as a function of the change in perimeter $\delta L$, or the corresponding rotation rate $\Omega$.

## V. READOUT PROCESS

In order to stabilize the perimeter of each CPE, we can send a master laser (Pr1/Pr2) to the side fiber in the CPE and detect the output signal using the detectors (D1/D2), as shown in Figure 3. The output $t_{Pr1}$ from the side fiber of Pr1 normalized by the input and that $t_{Pr2}$ of Pr2 can be calculated from

$$t_{Pr1} = t'_{CPE1} + \frac{r^2_{CPE1} t_{CPE2} t_{cav} g_{Br2} g_{Ab2}}{1 - t_{CPE1} t_{CPE2} t_{cav} g_{Br2} g_{Ab2}}, \quad (25)$$

$$t_{Pr2} = t'_{CPE2} + \frac{r^2_{CPE2} t_{CPE1} t_{cav} g_{Br1} g_{Ab1}}{1 - t_{CPE1} t_{CPE2} t_{cav} g_{Br1} g_{Ab1}}, \quad (26)$$

where $t_{CPEj}$ and $r_{CPEj}$ are the effective transmissivity and reflectivity of CPE$j$ ($j$=1,2) following Eqs. (1) and (2), $g_{Brj} = |g_{Brj}| e^{i\theta_{Brj}}$ is the Brillouin gain from P$j$ ($j$=1,2), and $g_{Abj} = |g_{Abj}| e^{i\theta_{Abj}}$ is the Brillouin absorption from P$j$ ($j$=1,2), and $t'_{CPEj}$ is determined by

$$t'_{CPEj} = \frac{\sqrt{1-k_{2j}} - \sqrt{1-k_{1j}} e^{i\phi_j}}{1 - e^{i\phi_j} \sqrt{1-k_{1j}} \sqrt{1-k_{2j}}}, \quad (27)$$

As the perimeter of the CPE$j$ ($j=1,2$) deviates from $L_{CPEj}$ by an amount of $\Delta L_{CPEj}$, the output changes as shown in Figure 7. Each of the output signals is sent to an AC-Servo, which stabilizes the perimeter of CPE$j$ to the center dip position where $\Delta L_{CPEj} = 0$ ($j=1,2$) as shown in Figure 7.

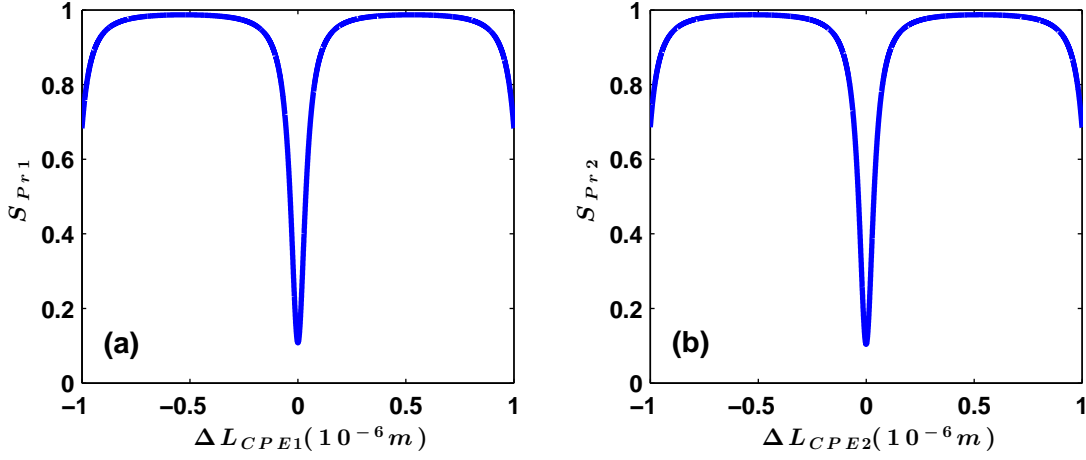

Figure 7 (a) Plot of the output signal $S_{Pr1} = |t_{Pr1}|^2$ versus the change $\Delta L_{CPE1}$ in the perimeter of CPE1; (b) plot of the output signal $S_{Pr2} = |t_{Pr2}|^2$ versus the change $\Delta L_{CPE2}$ in the perimeter of CPE2.

We explain below how we can observe the signals for the common mode effects and the differential effect. In the absence of rotation and acceleration, the frequency of the CW laser (B1) is assumed to be $f_{B1}$, and the frequency of the CCW laser (B2) is $f_{B2}$. The corresponding mode numbers are $m$ and ($m+12$), respectively. In the case we consider below, $m$ is around $2\times 10^7$.

The common mode effect can be represented as a common change in the perimeter $\delta L_{COM}$ for both B1 and B2. The frequency change in each Brillouin laser (B1/B2) as a result of the common mode effect is a function of $\delta L_{COM}$. Assume that the frequency change is $F_{COM1}(\delta L_{COM})$ for B1 and $F_{COM2}(\delta L_{COM})$ for B2. Similarly, the differential mode effect can be represented as a change in the perimeter $\delta L_{DIF}$ for B1 and $-\delta L_{DIF}$ for B2. The frequency change for each laser due to the differential mode effect is again a function of $\delta L_{DIF}$, which is assumed to be $F_{DIF1}(\delta L_{DIF})$ for B1 and $-F_{DIF2}(\delta L_{DIF})$ for B2, respectively. Notice that the functions $F_{COM,j}$ and $F_{DIF,j}$ ($j=1,2$) have odd symmetries. In the presence of possible rotation and acceleration, represented by $\delta L_{COM}$ and $\delta L_{DIF}$, we can express the lasing frequencies as

$$f'_{B1} = f_{B1} + F_{COM1}(\delta L_{COM}) + F_{DIF1}(\delta L_{DIF}) \tag{28}$$

$$f'_{B2} = f_{B2} + F_{COM2}(\delta L_{COM}) - F_{DIF2}(\delta L_{DIF}) \tag{29}$$

Since the mode number *m* is very large ($2\times 10^7$), the common-mode-effect induced frequency shifts have almost the same magnitudes in each direction, even though the base frequencies differ by 12 times the FSR. The same approximation also applies for the differential-mode-effect induced frequency shifts. We can then simplify the analysis by assuming $F_{COM1}(\delta L_{COM}) \cong F_{COM2}(\delta L_{COM}) \equiv F_{COM}(\delta L_{COM})$ and $F_{DIF1}(\delta L_{DIF}) \cong F_{DIF2}(\delta L_{DIF}) \equiv F_{DIF}(\delta L_{DIF})$.

The readout process measures the frequency differences induced in the two Brillouin lasers in order to extract the rotation and acceleration experienced by the system. There are four signals that are used in this process: the two Brillouin laser outputs (B1 and B2), and the pickoffs from the master lasers (Pr1 and Pr2).

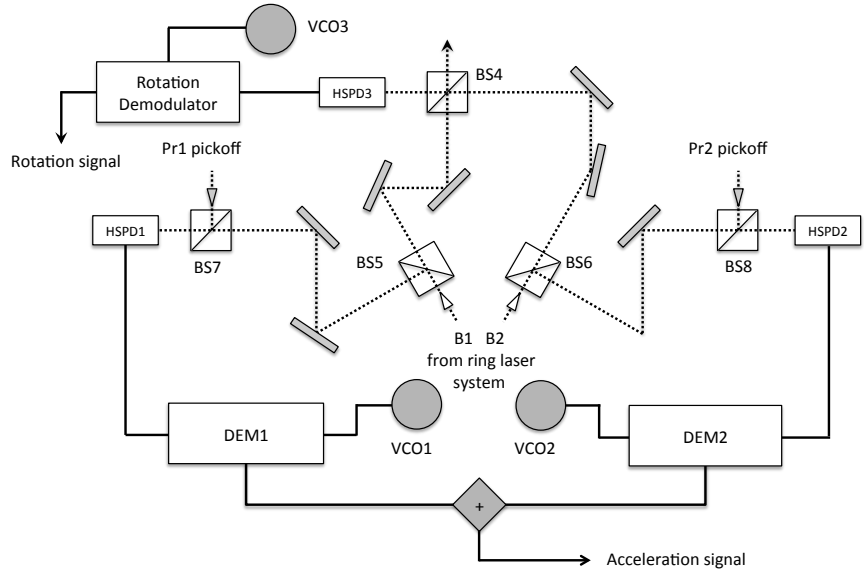

Figure 8 Schematic illustration of the readout section.

The two superluminal lasers, B1 and B2, are combined with each other on a 50/50 beamsplitter (BS4) and its output is directed into a high-speed photodetector (HSPD3). The beat frequency at HSPD3 will be $f_{B2} - f_{B1} - 2F_{DIF}(\delta L_{DIF})$, which is less (greater) than the frequency difference between the two superluminal lasers expected from an undisturbed system ($f_{B2} - f_{B1} \equiv \Delta f$) if $F_{DIF}(\delta L_{DIF})$ is positive (negative). The beatnote measured by this detector is then mixed with the output of a voltage controlled oscillator (VCO3) set to match $\Delta f$ and is demodulated using the rotation demodulator, which is a phase-locked-loop frequency demodulator (PLL-FD) [26]. As a result, we get a signal of $F_{DIF-DET} = |2F_{DIF}(\delta L_{DIF})|$. This gives the magnitude of rotation, but not direction [i.e., it does not reveal the sign of $F_{DIF}(\delta L_{DIF})$]. However, we can remedy this easily as follows. We modulate the frequency of VCO3 around its quiescent value of $\Delta f$. When the VCO3 frequency is $\Delta f + |\varepsilon|$, $F_{DIF-DET}$ will be greater than $|2F_{DIF}(\delta L_{DIF})|$ if the sign of $F_{DIF}(\delta L_{DIF})$ is positive, and less than $|2F_{DIF}(\delta L_{DIF})|$ if the sign of $F_{DIF}(\delta L_{DIF})$ is negative. Thus, we can determine both magnitude and sign of $F_{DIF}(\delta L_{DIF})$.

To determine the common mode effect, a small part of each Brillouin laser is combined with its corresponding master laser on a 50/50 beamsplitter (BS7/BS8) and the resulting beat signal is detected with a high-speed photodetector (HSPD1/HSPD2). The output of each detector is then mixed with the output of a voltage-controlled oscillator (VCO1/VCO2) set at $f_1/f_2$, which equals the difference in frequency between the master laser and the Brillouin laser from the undisturbed system. This signal is sent to a low pass filter so that we get only the difference frequency and then it is converted to a voltage signal using a PLL-FD (DEM1/DEM2). As a result, we get a DC signal proportional to the departure of each superluminal laser frequency from the value expected from an unmoving system.

Consider now the output of DEM1. It will produce a signal proportional to $\left|F_{COM}(\delta L_{COM}) + F_{DIF}(\delta L_{DIF})\right|$. By modulating VCO1 around $f_1$, we can determine the magnitude as well as the sign of the signal produced by beating the output of HSPD1 with VCO1. For example, by increasing the frequency of VCO4 slightly, if we see a decrease in the output, we can determine that $\left|F_{COM}(\delta L_{COM}) + F_{DIF}(\delta L_{DIF})\right| > 0$; otherwise, if we see an increase in the output, we can determine that $\left|F_{COM}(\delta L_{COM}) + F_{DIF}(\delta L_{DIF})\right| < 0$. Thus we can produce a voltage that corresponds to $\delta f_1 = F_{COM}(\delta L_{COM}) + F_{DIF}(\delta L_{DIF})$ rather than $\left|F_{com}(\delta L_{com}) + F_{dif}(\delta L_{dif})\right|$. Similarly, by modulating VCO2 around $f_2$, we can produce a voltage that corresponds to $\delta f_2 = F_{COM}(\delta L_{COM}) - F_{DIF}(\delta L_{DIF})$ rather than

$\left| F_{COM}(\delta L_{COM}) + F_{DIF}(\delta L_{DIF}) \right|$. Summing up these two signals yields $2F_{COM}(\delta L_{COM})$, from which we can determine the magnitude and sign of $F_{COM}(\delta L_{COM})$.

When the difference in rotational shift or acceleration-induced shift in each direction is taken into account, it is possible to account for this difference exactly as well. Now the output of DEM1 will be a signal proportional to $\left| F_{com1}(\delta L_{com}) + F_{dif1}(\delta L_{dif}) \right|$. Using the same method as above, we can determine the magnitude as well as the sign of the signal, i.e. we can determine $\delta f_1 = F_{com1}(\delta L_{com}) + F_{dif1}(\delta L_{dif})$. Similarly we can produce a voltage that corresponds to $\delta f_2 = F_{com2}(\delta L_{com}) - F_{dif2}(\delta L_{dif})$. Using the values of $\delta f_1$ and $\delta f_2$ and the knowledge of the functions $F_{com1}$ and $F_{com2}$, we can get the magnitude as well as the sign for both $\delta L_{com}$ and $\delta L_{dif}$, since we have only two unknowns and two equations.

## VI. CONCLUSION AND FUTURE WORK

We have developed an Active Fast Light Fiber Optics Sensor (AFLIFOS) that can be used for high sensitivity measurements of many parameters of interest, both due to common mode effects (e.g. acceleration, strain and temperature) and differential mode effect (e.g. rotation). The system has three main sections: the pump laser package, the ring resonator package and the readout system. The superluminal effects for two counter-propagating Brillouin laser modes are produced by two auxiliary resonators (CPE) that are coupled to the primary fiber resonator. We first developed a detailed theoretical model for optimizing the design of the AFLIFOS. We determined the enhancement factor of the sensitivity to be ~187 and ~−187 under the optimized condition, respectively for the two Brillouin lasers, when the "effective" change in perimeter of the primary fiber resonator is 0.1nm. By adjusting the parameters such as the perimeters and the coupling coefficients, it is possible to get much higher enhancement. We then showed the design of using master lasers for stabilizing the perimeter of the CPEs. We also showed the design for the readout section, where the beat between each superluminal laser and a reference master oscillator stabilized to an absolute atomic reference is measured. Assuming that the change in frequency for the lasers as a function of the change in the perimeter of the primary resonator is known, we can determine the common mode effects and differential mode effect separately.

In the future, we will carry out experimental demonstration of the AFLIFOS capable of measuring rotation as well as acceleration, strain, and temperature. Specifically, we will use a diode laser at ~1530 nm, locked to one of the optical resonances in an acetylene vapor cell using saturated absorption spectroscopy, and four different AOMs to produce two Brillouin pumps and two master lasers for stabilizing the auxiliary cavities. Commercially available and tunable couplers will be used to couple the primary resonator to the auxiliary cavities. Linear input-output sections of fibers will also be coupled to the primary resonator as well as the auxiliary cavities using the same approach. Once the system is constructed, we will use it to demonstrate measurement of rotation, acceleration, strain and temperature.

## VII.   ACKNOWLEDGEMENTS

This work has been supported by AFOSR grant FA9550-10-01-0228, NASA grants NNM13AA60C, NNX15CM35P and NNX16CM03C, and Darpa grant W911NF-15-1-0643.

TBD:

In general, gyroscopes require temperature controls. However, for the AFLIFOS we need to use active temperature controller to the precision of 1mK, because, as recently found in Ref. 27, Brillouin fiber lasers are strongly affected by thermal effects from the change in lasing power.